\documentclass[%
 reprint, 
 amsmath,amssymb,
 aps, physrev,
]{revtex4-2}

\usepackage[compat=1.1.0]{tikz-feynman}
\usepackage{subcaption} 
\usepackage{graphicx}
\usepackage{amsmath}
\usepackage{caption}    
\usepackage{tikz}

\usepackage{graphicx}
\usepackage{dcolumn}
\usepackage{bm}

\begin{document}

\preprint{APS/123-QED}

\title{\textbf{Event Rates at a 10~TeV Muon Collider and Implications for Detector Design: Trigger, Data Acquisition, and Luminosity} 
}%

\author{Tova Holmes and Lawrence Lee}
\affiliation{%
 University of Tennessee, Knoxville \\
 tholmes@utk.edu, llee@utk.edu
}%

\date{\today}

\begin{abstract}
This short document presents a discussion of rates of a wide range of processes at a muon collider. The goal is to provide a first look at what will fill the detector, how often processes of interest are occurring, and the implications for detector readout, a potential trigger system, and luminosity measurements.
\end{abstract}

\maketitle


\section{\label{sec:intro}Introduction}

A muon collider is an exciting possibility for the exploration of the energy frontier~\cite{Accettura:2023ked,AlAli:2021let,Black:2022cth}, which has received strong support in the United States as its next flagship particle physics project~\cite{P5:2023wyd, NationalAcademiesofSciencesEngineeringandMedicine:2025bix}. Experience with detector design at the Large Hadron Collider (LHC) tells us that several decades of dedicated R\&D are required before production, with especially long timelines for ASIC design. To reach the target of a mid-century muon collider, we must begin developing an understanding of its detector requirements today. 

This paper reviews our current understanding of what signals and backgrounds a muon collider detector will experience, and discusses its implications, working within the context of the two 10 TeV detector designs, MAIA and MUSIC ~\cite{bell2025maianewdetectorconcept, accettura2025muoncollider}, creates so far.

\section{\label{sec:rates}What Physics Occupies a Muon Collider Detector?}

By far the dominant process in a muon collider detector is Beam-Induced Background (BIB) from the decays of beam muons. Current detector designs include tungsten nozzles with additional layers of borated polyethelene designed to shower the electrons emerging from beam muon decays and absorb the majority of the energy~\cite{Accettura:2023ked,Ally:2022rgk}. The resulting BIB is high-multiplicity, low-energy, and emerging along the surfaces of the nozzles, concentrated at their tips~\cite{accettura2025muoncollider}. Compared to pile-up processes at a $pp$ collider, BIB properties are much more distinct from signal, making it easier to subtract as long as full reconstruction can be performed.

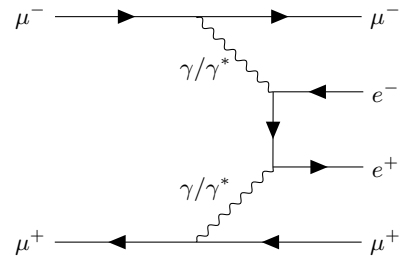
\begin{figure}[htbp]
    \centering
    \begin{tikzpicture}
    \begin{feynman}
      \vertex (i1) {\( \mu^- \)};
      \vertex [below=3cm of i1] (i2) {\( \mu^+ \)};
      \vertex [right=2.2cm of i1] (v1);
      \vertex [right=2.2cm of i2] (v2);
      \vertex [right=2.2cm of v1] (f1) {\( \mu^- \)};
      \vertex [right=2.2cm of v2] (f2) {\( \mu^+ \)};
      \vertex [below=1cm of f1] (e1) {\( e^- \)};
      \vertex [above=1cm of f2] (e2) {\( e^+ \)};
      \vertex [left=1.5cm of e1] (v3);
      \vertex [left=1.5cm of e2] (v4);

      \diagram* {
        (i1) -- [fermion] (v1) -- [fermion] (f1),
        (i2) -- [anti fermion] (v2) -- [anti fermion] (f2),
        (v1) -- [photon, edge label'=\( \gamma/\gamma^* \), inner sep=1pt] (v3),
        (v2) -- [photon, edge label=\( \gamma/\gamma^* \), inner sep=1pt] (v4),
        (e1) -- [fermion] (v3) -- [fermion] (v4) -- [fermion] (e2),
      };
    \end{feynman}
    \end{tikzpicture}
    \label{fig:landau-lifshitz}
    
  \caption{Example processes for incoherent $e^+e^-$ scattering. Photons can be real or virtual. Scattering can also occur between one real and one virtual photon. In any of these cases, the two photons emerge from the two beams.}
  \label{fig:incoherent-scattering}
\end{figure}

After BIB, the next highest rate processes are from particle production at the interaction point (IP) itself. For example, low energy electron-positron pairs can be produced from beam interactions like the ones shown in Figure~\ref{fig:incoherent-scattering}. Similar processes will also lead to hadron pair production with a large rate, and at large energies a significant fraction comes from the quark and gluon contributions to the muon PDF~\cite{Han:2021kes}. The electron-positron pair production is steeply falling in momentum. The hadrons are similarly produced with a steeply falling momentum distribution and can lead to detector signals, but the rate of producing GeV-scale physics is suppressed by orders of magnitude.

A high magnetic field curls most of these particles near the beamline, preventing them from reaching active detector elements, but the high energy tail can enter the detector and increase occupancy in the innermost layers. 10~TeV detector designs use a 5~T solenoidal magnet within the tracking region. While giving transverse momentum resolution to more energetic particles, increasing this field also helps to control this low-momentum background. To reach the innermost layer of the vertex detector (at a radius of 3 cm in the MAIA and MUSIC detectors), charged particles must have a minimum transverse momentum of 23~MeV. To traverse the full tracker, 1.5 m in radius for both detectors, a minimum of 1.1 GeV is required.

\begin{figure*}[p]
    \centering
    \includegraphics[width=0.85\linewidth]{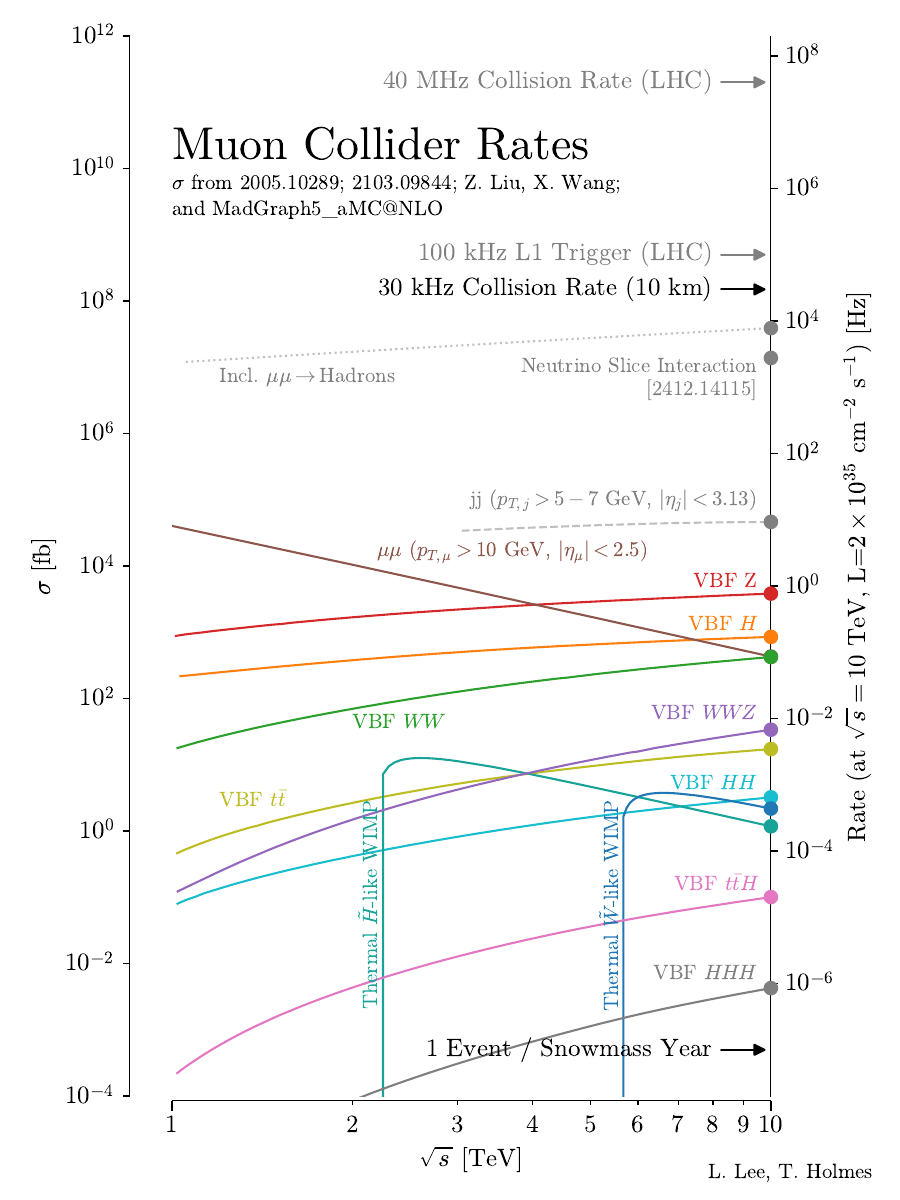}
    \caption{Process cross-sections at a muon collider with $\sqrt{s}$ ranging from 1 to 10 TeV. On the right-hand axis, the expected 10 TeV instantaneous luminosity~\cite{Accettura:2023ked} is used to translate these cross-sections into rates in the detector. Background-like processes are shown in light grey, while processes of interest are in solid lines of various colors. Cross section values are from various references~\cite{Bojorquez-Lopez:2024bsr,LiuWang,Han:2021kes,Costantini:2020stv}  and the $\mu\mu$ final state cross sections were calculated directly using MadGraph5\_aMC@NLO~\cite{Alwall:2014hca}.}
    \label{fig:rates}
\end{figure*}

While the majority of BIB is due to the electrons resulting from muon decays, neutrinos are produced with such a high intensity that their interactions must also be considered. For the most part, neutrinos produced in these decays will propagate in straight lines, unaffected by magnetic fields or materials of the machine. First studies of their interactions were performed for several muon beam scenarios, with the conclusion that neutrinos would interact with a $\sqrt{s}=$~10 TeV muon collider detector region in 44\% of all bunch crossings~\cite{Bojorquez-Lopez:2024bsr}. These neutrino interactions are concentrated near the negative $x$ axis of the detector, \emph{i.e.} at an azimuthal angle near $\pi$, and most of these interactions occur in the nozzles, but can lead to showers into the sensitive detector region. 21\% of these interactions will occur in the active detector itself, leading to energy depositions at the 100 GeV to TeV scale.~\cite{Bojorquez-Lopez:2024bsr}

Hard-scattering processes of interest appear at much lower rates than these low-energy processes. When the cross section values are interpreted in an instantaneous luminosity scenario of $L=2\times10^{35}$~cm$^{-2}$~s$^{-1}$ at $\sqrt{s}=10$~TeV~\cite{Accettura:2023ked}, these processes begin occurring at around $10^{-4}$ times the bunch crossing rate of $30$~kHz (for a collider ring of 10~km). The vast majority of collision events will not involve physics processes of interest for this machine. For example, VBF production of a Z boson, with a cross section of 3900 fb, will occur with a rate of 0.77 Hz. This is roughly one in every 40,000 bunch crossings. VBF production of one and two Higgs boson will occur at rates of 0.17 Hz and 0.65 mHz, respectively. A summary of these values can be found in Figure~\ref{fig:rates}. This relative rate may have significant implications for the design of the detector system and physics analysis optimization, as discussed below.

\section{Implications for Detector Readout}

Early experimental considerations for energy frontier muon colliders have been focused on triggerless detector readout. Many of these thoughts have led to significant on-detector BIB filtering and data compression to enable a full-detector readout of every collision. However, the narrative of Figure~\ref{fig:rates} shows that a triggerless system would require a performant readout system to largely read out soft background processes.

The possibilities for a trigger system require in-depth study. Tracking is the most reliable BIB suppression technique, so it is likely that any viable trigger system would require a method for real-time tracking. With a requirement of a prompt track at the GeV scale alone, the event rate could be reduced by several orders of magnitude, assuming the fake rate could be kept low. However, maintaining inclusivity to unconventional signatures with this strategy is not a given. Flagship analyses, like the disappearing track search that would give sensitivity to the thermal WIMP scenarios shown in Figure~\ref{fig:rates}, may be left out of such a trigger strategy. Careful review of background properties, detector capabilities, and signals of interest is required to understand the optimal balance between event selection and simplified event read-out. It is unlikely that an LHC-style triggering system would be necessary, but a simplified, inclusive near-trigger, operating locally within subsets of the detector may be advantageous.


In collider detectors, a driving element of tracker performance is the overall amount of material required. This tracker mass is often set by the needed cooling systems, which in turn, originate from the detector's power requirements. As we explore smarter electronics for the reduction of the BIB, this will naturally lead to larger power consumption from our on-board electronics. However, a large fraction of the power consumption is from the transmission of data off detector. A reduction of the full readout rate may reduce the transmission power, allowing for more power-hungry algorithms to be run locally on the electronics of the detector.

Even with these fractional considerations, the absolute rates discussed here are well below those needed at today's LHC. A 30 kHz collision rate is already far below the LHC experiments's Level-1 readout rate of about 100~kHz, and far below those of the HL-LHC experiments that push to the MHz level. This implies that as we scale up the complexity of the detector and its readout system, the potential allowed latencies are large compared to contemporary systems.

\section{Implications for Luminosity Measurement}

\begin{figure*}
    \centering
    \includegraphics[width=0.9\linewidth]{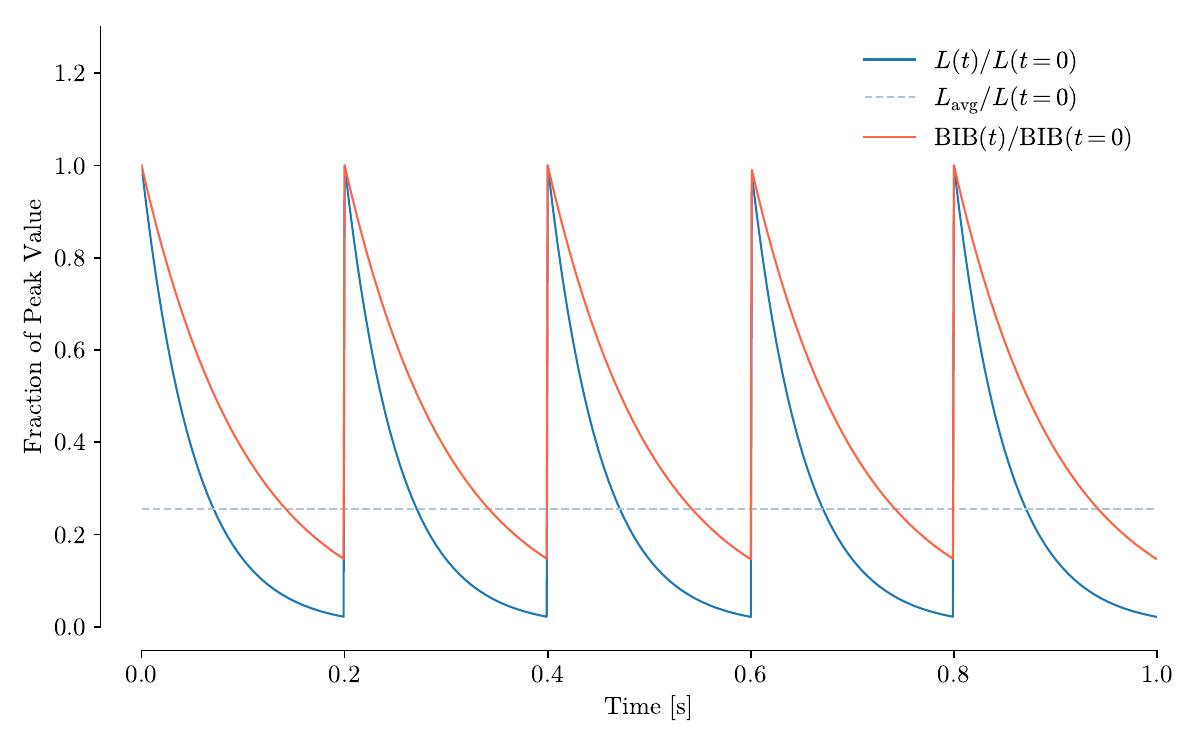}
    \caption{Comparison of variation in instantaneous luminosity and detector BIB on the time scales of bunch injection. Both are calculated using $\sqrt{s} = $ 10 TeV and an injection rate of 5 Hz. BIB is proportional to the number of muons in the ring, while instantaneous luminosity is proportional to the square of number of muons, assuming other beam parameters are fixed.}
    \label{fig:instantaneous_luminosity}
\end{figure*}

The luminosity at a muon collider varies substantially at the scale of the repetition rate, the rate at which muons are reinjected into the collider ring. For a $\sqrt{s} =$ 10 TeV collider, the 5 TeV muons have an average lab-frame lifetime of 0.10 s. The repetition rate is chosen based on this lifetime, typically taken to be 5 Hz for a 10 TeV collider. Figure~\ref{fig:instantaneous_luminosity} shows the impact of this variation in beam intensity on both luminosity and BIB intensity. Because of their correlation, most processes of interest will occur in the highest BIB events. In addition, this rapidly changing BIB occupancy means that any on-detector mitigation strategy will need to be able to accommodate the constant variation.

The most common process in the detector, the BIB, cannot be used for a luminosity measurement. Because these particles originate from decays in single beams, the BIB is proportional to the total current in each beam. The luminosity of the collisions, in addition to relying on this current, also depends on the beam emittance and beam profile, not accessible from measurements of the BIB. For this same reason, LHC strategies for luminosity measurement, which in most cases rely on very forward detectors, are challenging in the presence of the high rates of BIB, which make instrumentation challenging and provide overwhelming backgrounds for a measurement of luminosity.

The incoherent $e^+e^-$ scattering is the result of interactions between the two beams. It is possible that using occupancy in the innermost layers of the detector, or the higher-$p_\mathrm{T}$, lower-rate tail that travels through more layers of the detector, might yield a luminosity measurement. However, there will be a large contribution from BIB that would need to be measured and controlled in order for these rates to be used. The pure detector contribution from BIB could be measured by relaxing the final beam focus, reducing the collision luminosity while leaving the beam current the same.

Previous work studied the possibility of using high-angle $\mu$-Bhabha scattering to measure luminosity at a $\sqrt{s} =$ 1.5 TeV muon collider, and estimated that this could yield 0.1\% accuracy~\cite{Giraldin:2021gxz}. However, as shown in Figure~\ref{fig:rates}, this process decreases sharply with $\sqrt{s}$, and becomes subdominant to processes like VBF $Z$ production above $\sqrt{s} = $ 4 TeV. Above these energies, simple $Z$ counting or measurements of the background processes may be more effective, but additional studies are needed to explore other useful physics processes or methods.

\section{Conclusion}

There is much still to be understood about the optimal design of a muon collider detector. We are only beginning to understand the realities of background processes and event reconstruction at 10 TeV. With this basic understanding forming the foundation, now is the time to begin understanding the realities of operating these detectors, reading them out, and extracting all of the physics that a muon collider can produce.

\vspace{1em}

\begin{acknowledgments}
For invaluable discussions, we thank Kiley Kennedy, Zhen Liu, Ben Rosser, Angira Rastogi, Massimo Casarsa, Sergo Jindariani, Federico Meloni, Simone Pagan Griso, Isobel Ojalvo, Tao Han, Matheus Hostert, and Xing Wang
in helping to helped sculpt these thoughts, plots, and arguments. We thank the Galileo Galilei Institute for Theoretical Physics providing a venue for many of the discussions that contributed to this work. This work was supported by funding from the Department of Energy Office of Science, award numbers DE-SC0020267 and DE-SC0023122; the National Science Foundation, under award number 2235028; and the Research Corporation for Science Advancement. 
\end{acknowledgments}

\bibliography{apssamp}

\end{document}